%% file: main.tex
\documentclass[12pt]{iopart}

\usepackage{iopams}  
\usepackage{graphicx}
\usepackage{cite}
\usepackage{hyperref}
\usepackage[dvipsnames]{xcolor}

\expandafter\let\csname equation*\endcsname\relax
\expandafter\let\csname endequation*\endcsname\relax
\usepackage{amsmath}

\begin{document}

\title[]{Ultracold $^{88}\rm{Sr}_2$ molecules in the absolute ground state}

\author{K. H. Leung$^{1,*}$, E. Tiberi$^{1}$, B. Iritani$^{1}$, I. Majewska$^{2}$, R. Moszynski$^{2}$, and T. Zelevinsky$^{1,\dag}$}

\address{$^1$ Department of Physics, Columbia University, 538 West 120th Street, New York, NY 10027-5255, USA}
\address{$^2$ Quantum Chemistry Laboratory, Department of Chemistry, University of Warsaw, Pasteura 1, 02-093 Warsaw, Poland}

\begin{abstract}
We report efficient all-optical creation of an ultracold gas of alkaline-earth-metal dimers, $^{88}\rm{Sr}_2$, in their absolute ground state. Starting with weakly bound singlet molecules formed by narrow-line photoassociation in an optical lattice, followed by stimulated Raman adiabatic passage (STIRAP) via a singlet-dominant channel in the $(1)0_u^+$ excited potential, we prepare pure samples of more than 5500 molecules in $X^1\Sigma_g^+(v=0,J=0)$.  We observe two-body collisional loss rates close to the universal limit for both the least bound and most bound vibrational states in $X^1\Sigma_g^+$.  We demonstrate the enhancement of STIRAP efficiency in a magic-wavelength optical lattice where thermal decoherence is eliminated.  Our results pave the way for the use of alkaline-earth-metal dimers for high-precision spectroscopy, and indicate favorable prospects for robust quantum state preparation of ultracold molecules involving closed-shell atoms, as well as molecule assembly in deep optical traps tuned to a magic wavelength.
\end{abstract}

%

\vspace{2pc}
\newline\noindent
%
%
%


$^*$ kl2908@columbia.edu

$^\dag$ tanya.zelevinsky@columbia.edu

\section{Introduction}

Molecules serve as natural test beds for molecular quantum electrodynamics \cite{alighanbari2020precise,lai2019,patra2020proton,kortunov2021proton}, in addition to tests of beyond-Standard-Model physics such as searches for $T$-symmetry violation \cite{Andreev2018,Kozyryev2017,denis2019,yu2021,ho2020new,grasdijk2021centrex}, dark matter \cite{Kozyryev2021,Roussy2021,antypas2021probing}, time variation of fundamental constants \cite{Kobayashi2019,truppe2013search,bagdonaite2013stringent,Shelkovnikov2008,Hudson2006cold}, and non-Newtonian gravity \cite{Borkowski2019}. In this regard, alkaline-earth-metal molecules represent an exciting frontier since their closed-shell structure lead to $^1\Sigma$ ground potentials that are naturally insensitive to external perturbations. Furthermore, homonuclear combinations possess high-$Q$ subradiant transitions to the excited potentials that can be utilized as \textit{in situ} magnetometers for systematic calibrations. These qualities present a pristine environment for precise molecular spectroscopy as a means to probe new physics. In addition, molecular gases at cold or ultracold temperatures provide experimental benchmarks for understanding collisions and chemical reactions, where accurate theoretical descriptions that include quantum effects remain a significant computational challenge \cite{heazlewood2021,Krems2008,Jurgilas2021,liu2021precision,hu2021nuclear,wolf2017state,segev2019collisions,sawyer2011cold,guo2018pra,Guo2018prx}. In these applications, often it is necessary to perform tailored state preparation and measurement of the molecules in specific rovibronic and hyperfine states.

On another front, the ability to prepare molecular ensembles with increasingly greater phase-space densities has been driven by efforts to observe novel many-body physics \cite{Micheli2006,Yi2007,Capogrosso2010,gadway2016strongly,Blackmore2018} and to realize platforms for quantum information processing \cite{Ni2018,Hudson2018,Yu2019,sawant2020ultracold,lin2020quantum,albert2020,campbell2020,Hughes2019,Demille2002} that take advantage of the rich rovibronic structure of molecules. In one approach to the creation of ultracold molecular ensembles, molecules are associated via optical or magnetic Feshbach resonances from laser-cooled ultracold atoms. By preserving the original atomic phase-space density, this has led to the creation of degenerate Bose \cite{greiner2003emergence,jochim2003bose,zwierlein2003observation} and Fermi \cite{de2019degenerate} alkali-metal diatomic molecular gases. However, in many practical applications, it is advantageous to initialize the molecules in a deeply bound state. For example, in bi-alkalis \cite{danzl2008quantum,ni2008high,aikawa2010coherent,danzl2010ultracold,takekoshi2014ultracold,molony2014,Park2015,Guo2016,Rvachov2017,seeselberg2018,Liu2019Obs,Voges2020}, a common step involves de-exciting the weakly bound Feshbach molecules to the absolute ground state using stimulated Raman adiabatic passage (STIRAP) in order to access the large molecule-frame dipole moments. Despite these technical successes \cite{bergmann2019roadmap} and prior work on weakly-bound molecules of $\rm{Sr}_2$ \cite{ciameipra2017,Stellmer2012prl,Reinaudi2012} and $\rm{Yb}_2$ \cite{borkowski2017prabeyond}, the production of ultracold molecules consisting of closed-shell atoms in the absolute ground state has yet to be demonstrated. 

In this paper, we extend the robust state control offered by STIRAP to the entire ground potential of $^{88}\rm{Sr}_2$, a homonuclear alkaline-earth-metal molecule. This paper is organized into three main parts. First, we describe the spectroscopy and transition strength measurements of the $(1)0_u^+$ and $X^1\Sigma_g^+$ potentials in order to identify a feasible STIRAP pathway for adiabatic transfer within $X^1\Sigma_g^+$. Next, as proof of concept, we experimentally perform STIRAP to create absolute-ground-state molecules and investigate factors that limit the transfer efficiencies both in free flight and in a magic-wavelength optical lattice. Finally, we study the lifetime of the absolute-ground-state molecules and measure the two-body inelastic loss rates to extend our understanding of ultracold collisions, which may potentially inform the feasibility of producing stable molecular Bose-Einstein condensates \cite{ciamei2017observation,heinzen2000superchemistry,stellmer2013pra}. We expect that the techniques described herein can be generalized to molecules consisting of at least one closed-shell atom \cite{barbe2018observation,roy2016pra,Green2020prx,green2019pra,Guttridge2018two,yangpra2019magnetic}, and to other molecular species in deep optical traps such as tweezers and lattices \cite{Cairncross2021,he2020coherently,anderegg2019optical,langin2021polarization,brooks2021preparation,guan2021,wu2021high}.

\section{Experimental methods}

We employ a standard two-stage magneto-optical trap (MOT) cycling on the broad $^1S_0$-$^1P_1$ and narrow $^1S_0$-$^3P_1$ transitions in $^{88}\mathrm{Sr}$, while repumping on $^3P_2$-$^3S_1$ and $^3P_0$-$^3S_1$. This prepares the atoms in the $^1S_0$ electronic ground state at approximately $2(1)\,\mu \mathrm{K}$ as inferred from a time-of-flight ballistic expansion. During the narrow-line MOT cooling stage, the atoms are overlapped with a one-dimensional optical lattice with a typical trap depth of $50\,\mu \mathrm{K}$ where they remain optically trapped after the MOT coils are ramped off. In this work, the light for the lattice is derived from a Ti:sapphire laser. To produce weakly bound molecules in the electronic ground potential, $X^1\Sigma_g^+$, we photoassociate the atomic sample with a 1.5 ms laser pulse to $(1)0_u^+(v=-4,J=1)$ where a sizable fraction subsequently decays down to the least bound vibrational states $X(62,0)$ and $X(62,2)$ \cite{Reinaudi2012}. In a few instances, we alternatively photoassociate to $(1)0_u^+(-5,1)$ to produce $X(61,0)$ and $X(61,2)$. 

The nuclear spin of $^{88}\mathrm{Sr}$ is $I=0$, resulting in the absence of hyperfine structure. Spin statistics forces the molecular wavefunction to be even upon the exchange of the bosonic nuclei in $^{88}\mathrm{Sr}_2$. This implies that only even $J$ (total angular momentum of the molecule) are allowed in the gerade $X^1\Sigma_g^+$, and only odd $J$ are allowed in the ungerade $(1)0_u^+$. Since the total number of bound states in a given potential is not always known \textit{a priori}, in some cases we use negative values for $v$ (the vibrational quantum number) that count down from the dissociation threshold of the respective potentials.

While we typically work with nearly equal mixtures of $J=0,2$ ground state molecules, we can perform a purification step that clears away $J=2$ molecules. This is done by photodissociating $J=2$ molecules to the ${^1S}_0+{^3P}_1$ threshold with an additional laser for 1 ms, and blasting the resulting atoms out of the trap with 461 nm laser light near-resonant with $^1S_0$-$^1P_1$. In either case, we prepare $\sim10^4$ molecules in $J=0$ to serve as our spectroscopic signal. We absorption-image the atomic fragments on $^1S_0$-$^1P_1$ after photodissociating $J=0$ molecules above the ${^1S}_0+{^3P}_1$ threshold with a 120 $\mu s$ pulse. While we usually perform absorption imaging at a slight grazing angle along the lattice direction to maximize the optical depth (and hence the signal-to-noise ratio), for calibration purposes we also take images perpendicular to the lattice direction from which we deduce that the molecules typically fill $\sim$570 lattice sites.

We use diode lasers for both the pump and anti-Stokes STIRAP lasers. The pump laser is stabilized using the Pound-Drever-Hall technique to a high-finesse cavity ($\mathcal{F}>10^5$, ultralow expansion glass). This phase stability is transferred to the frequency comb spectrum of a low-noise erbium-doped femtosecond fiber laser (FC1500-250-ULN, Menlo Systems GmbH) by referencing the repetition rate to the pump laser. By phase locking the anti-Stokes to the comb, the relative frequency stability of the pump and anti-Stokes is maintained to ${<1}$ kHz (estimated from initial scans of the Raman clock transition $X(62,2)\rightarrow X(0,0)$ using these two lasers; the true value is expected to be at the Hz level). The STIRAP lasers pass through a common Glan-Thompson polarizer with a 100,000:1 extinction ratio to drive $\pi$-transitions, and co-propagate along the lattice to minimize momentum transfer.

\section{High-resolution spectroscopy}

\begin{figure}
    \centering
    \includegraphics[width=\textwidth]{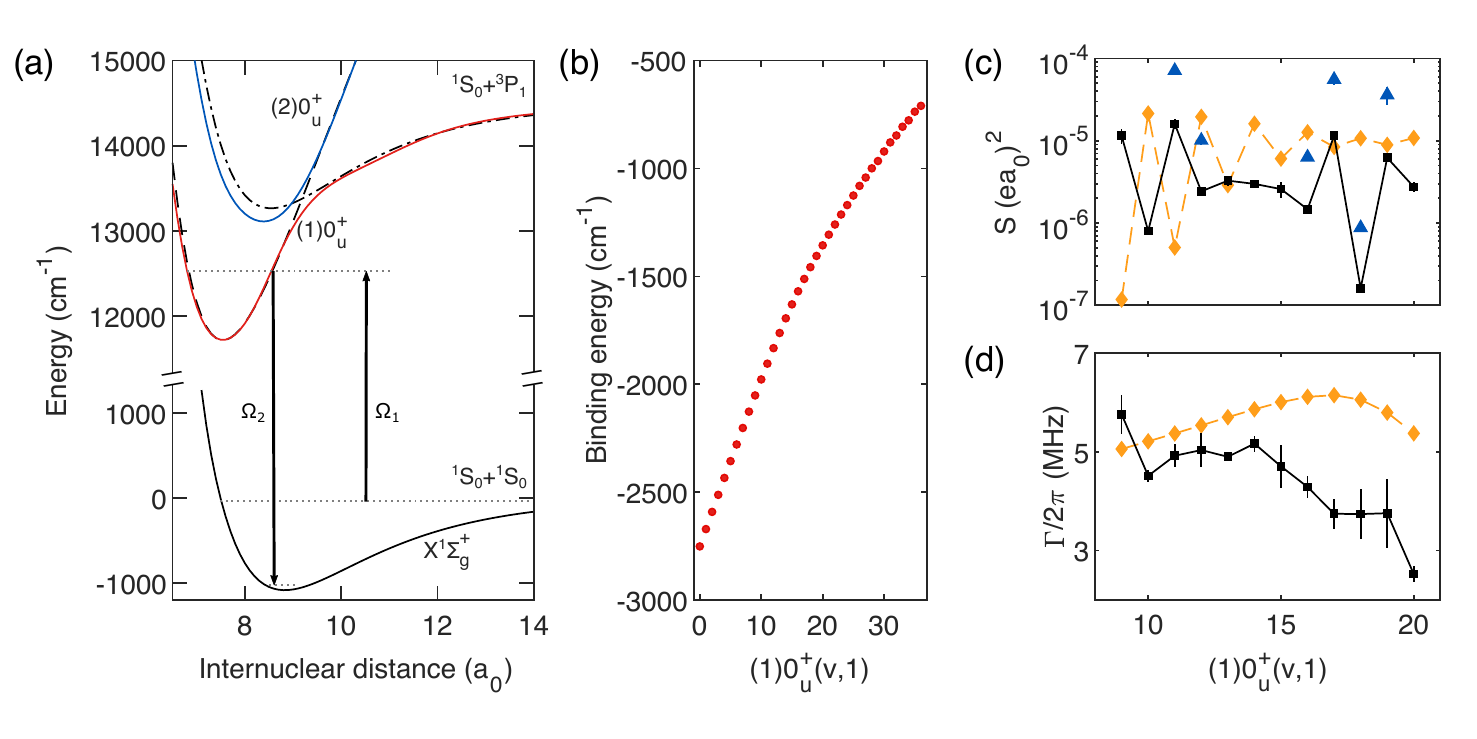}
    \caption{One-photon spectroscopy of $(1)0_u^+$. (a) The spin-orbit interaction couples $A^1\Sigma_u^+$ and $c^3\Pi_u$ (dashed and dotted-dashed respectively) in the manifold of excited potentials resulting in the admixed relativistic potentials $(1)0_u^+$ (solid red) and $(2)0_u^+$ (solid blue). Low-lying states of $(1)0_u^+$ present a pathway for adiabatic transfer in the ground potential $X{^1\Sigma}_g^+$ (solid black). $\Omega_1$ and $\Omega_2$ are the angular Rabi frequencies of the pump and anti-Stokes lasers respectively. For the spectroscopy of $(1)0_u^+$, only the pump laser is present. (b) Binding energies of the first 37 vibrational levels belonging to $(1)0_u^+$ with $J=1$. The marked change in trend at approximately $-1500\,\rm{cm}^{-1}$ can be attributed to the avoided crossing of the non-relativistic potentials. (c) Transition strengths $S$ to $(1)0_u^+(v=9-20,1)$ from $X(62,0)$ (black squares) and $X(61,0)$ (blue triangles). (d) Measured linewidths of the $(1)0_u^+$ states (black squares). Yellow diamonds are the theory predictions of $S$ and $\Gamma$ using the Morse/long-range $(1)0_u^+$ potential described in reference \cite{leung2020}. All error bars represent $1\sigma$ uncertainties. The binding energies and transition strength values are listed in table \ref{tab:0ubindtable} in the Appendix.}
    \label{fig:0uspec}
\end{figure}

Just as for atomic Sr, due to the heavy mass of the strontium nucleus the singlet and triplet electronic states of $\rm{Sr}_2$ are strongly mixed by the spin-orbit interaction \cite{Skomorowski2012jcp}. This effect is manifest in the relativistic potential $(1)0_u^+$ that asymptotes to the ${^1S}_0+{^3P}_1$ threshold, which can be identified as the lower branch of the avoided crossing between the $A^1\Sigma_u^+$ and $c^3\Pi_u$ non-relativistic potentials (figure \ref{fig:0uspec}(a)). We spectroscopically locate the first 37 vibrational levels of $(1)0_u^+$ with $J=1$ via one-photon excitation of $X(62,0)$. Previous studies have hitherto only probed weakly bound states near the intercombination \cite{Borkowski2014,reschovsky2018narrowline,Zelevinsky2006,McGuyerNJP2015,McGuyer2015control,mcguyerzelevinsky2015,McDonald2017}. As shown in figure \ref{fig:0uspec}(b), the effect of the spin-orbit perturbation is markedly noticeable from the change in the trend of the binding energies near $-1500 \,\rm{cm}^{-1}$ with respect to the ${^1S}_0+{^3P}_1$ threshold. Having found the rovibrational ground state of $(1)0_u^+$, we fit the binding energies of the first 11 states to the energies of a vibrating-rotor (valid only for low-lying states), \begin{equation}\label{eq:vibrot}
    E(v,J) = -D_e + \omega_e\left(v+\frac{1}{2}\right) - \omega_ex_e\left(v+\frac{1}{2}\right)^2 + \left[B_e-\alpha_e\left(v+\frac{1}{2}\right)\right]J(J+1)
\end{equation} where $\omega_e$, $x_e$, $B_e$, $\alpha_e$, and $D_e$ are the vibrational, anharmonicity, rotational, and vibration-rotation coupling spectroscopic constants and the potential depth, respectively. Our results, listed in table \ref{tab:0uspecconst}, compares favorably with those quoted for $A^1\Sigma_u^+$ inferred from the observed spectra of several isotopic combinations in reference \cite{Stein2011}. 
\begin{table}
\caption{\label{tab:0uspecconst} Extracted spectroscopic constants of $(1)0_u^+$ in units of $\rm{cm}^{-1}$. These were obtained from fitting the 11 lowest bound states to the energies of a vibrating rotor. Since we only measured states of the same angular momentum ($J=1$), we cannot separate out the vibrational and rotational constants.}
\begin{center}
\begin{tabular}{ccc}
\br
Spectroscopic constant& This work & Reference \cite{Stein2011}\\
\mr
$-D_e+2B_e$&-2791.21(11)&-2790.898\\
$\omega_e-2\alpha_e$&81.032(47)&80.713\\
$\omega_ex_e$&0.3327(42)&0.2296\\
\br
\end{tabular}
\end{center}
\end{table}

For coherent transfer within the singlet $X^1\Sigma_g^+$ ground potential, we ideally require intermediate states with marginal triplet admixture, favorable Frank-Condon overlap with the initial and final states, and a narrow linewidth. In an earlier work \cite{Skomorowski2012pra}, we predict that the first two criteria can be satisfied in the vicinity of the $A^1\Sigma_u^+$-$c^3\Pi_u$ avoided crossing, where transitions strengths with mid-to-low-lying states of $X^1\Sigma_g^+$ are expected to be as large as $10^{-2} (ea_0)^2$, while simultaneously maintaining reasonable transition strengths with photoassociated weakly bound molecules. As shown in figures \ref{fig:0uspec}(c) and (d), we verify this by measuring the transition strengths of $X(62,0)$ and $X(61,0)$ to $(1)0_u^+$ states on both sides of the avoided crossing, as well as their linewidths, $\Gamma$. To do so, we follow references \cite{guo2017praSPECTROSCOPY,debatin2011molecular} and measure the number of remaining molecules after a depletion pulse, $N(\delta_1,t)$, as a function of the pulse time $t$ and angular frequency detuning $\delta_1$ of the pump laser with respect to the ${X\rightarrow 0_u^+}$ transition. Straightforward rate equations imply that \begin{equation}
    N(\delta_1,t) = N_0 \exp{\left[-t\,\Omega_1^2\frac{\Gamma}{\Gamma^2 + 4\delta_1^2}\right]}, \label{eq:depletion}
\end{equation} where $N_0$ is the initial molecule number and $\Omega_1$ is the angular Rabi frequency of the ${X\rightarrow 0_u^+}$ pump transition. By simultaneously fitting the depletion curves versus $t$ and $\delta_1$, we extract $\Omega_1$ and $\Gamma$. We then convert $\Omega_1$ into transition strength, $S$, from conservative estimates of the pump laser beam waist and power. 

Our data suggests that $(1)0_u^+(11,1)$ offers one of the strongest pump couplings. Moreover, the required laser wavelengths of 793 nm and 732 nm (for the pump and anti-Stokes respectively) to address the entire depth of $X^1\Sigma_g^+$ are within the operating range of commercially available AR-coated laser diodes and our frequency comb. While transitions starting from $X(62,0)$ are generally weaker than those from $X(61,0)$, this is outweighed by the fact that we create larger samples of the former and detect them with better efficiency. Thus, in this study we choose $X(62,0)$ as our initial state and the singlet-dominant $(1)0_u^+(11,1)$ as the intermediate state for the STIRAP transfer. We note that a similar singlet pathway has been recently shown to be favorable for bi-alkali molecules \cite{yang2020}.

\begin{figure}
    \centering
    \includegraphics[width=\textwidth]{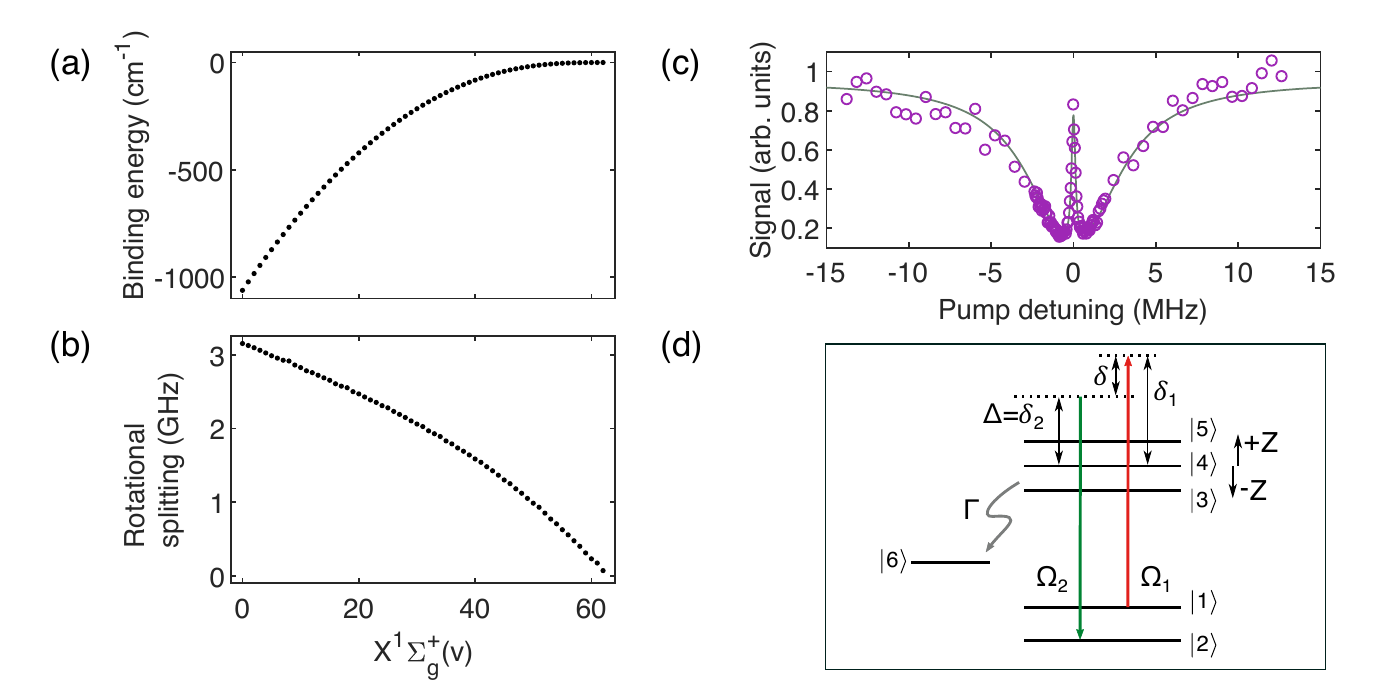}
    \caption{Two-photon Raman spectroscopy of ${X^1\Sigma}_g^+$. Here, both the pump and anti-Stokes lasers address molecular states. (a) Binding energies of all 63 vibrational levels in ${X^1\Sigma}_g^+$ ($J=0$), with respect to the ${^1S}_0+{^1S}_0$ threshold. (b) Rotational splittings between $J=2$ and $J=0$ ground state molecules. Error bars are smaller than the symbol size. (c) EIT spectrum in a $\Lambda$-system formed by $X(62,0)$, $X(0,0)$ and $(1)0_u^+(11,1)$. Each point is a single experimental shot. Solid line is the fit to the data using the analytical expression in equation \ref{eq:eitfit}. The only free parameters are $\delta_2$ and $\Gamma_{\mathrm{eff}}$, since $\delta_1$, $\Omega_1$, and $\Omega_2$ can be independently obtained from spectroscopy. (d) Level diagram of the quantum states in the numerical model described in the text.}
    \label{fig:Xspec}
\end{figure}

To locate the bound states of $X^1\Sigma_g^+$, we perform dark-resonance spectroscopy. We lock the pump laser on resonance with $X(62,0)\rightarrow (1)0_u^+(11,1)$ and adjust the laser power and pulse time such that we achieve nearly full depletion without bleaching the spectroscopic signal. For this part of the study, we use a tunable Ti:Sapphire laser as the anti-Stokes ($\sim$500 $\rm{W\,cm^{-2}}$). On resonance, $(1)0_u^+(11,1)$ is significantly Stark shifted by the anti-Stokes and the spectroscopic signal is no longer depleted by the pump. In this way, we find all 63 vibrational states in $X^1\Sigma_g^+$ with $J=0$ and $2$, as shown in figures \ref{fig:Xspec}(a) and (b). To compare our spectroscopy with Fourier transform spectra in a heatpipe from older studies \cite{Stein2008,Stein2010}, we note from equation \ref{eq:vibrot} that \begin{equation} \label{eq:rotsplit}
    E(v,2) - E(v,0) = 6\left[B_e-\alpha_e\left(v+\frac{1}{2}\right)\right].
\end{equation} By simultaneously fitting the binding energies and the $J=0,2$ rotational splittings of the first 5 vibrational levels to equations \ref{eq:vibrot} and \ref{eq:rotsplit} respectively, we extract the spectroscopic constants (see table \ref{tab:Xspecconst}) and find excellent agreement with the values reported in references \cite{Stein2008,Stein2010} that are weighted across several isotopologues. To verify that the addressed states form a $\Lambda$-system, we lock the anti-Stokes laser on resonance with the absolute rovibrational ground state of the $^{88}\rm{Sr}_2$ dimer (i.e. $X(0,0)\rightarrow (1)0_u^+(11,1)$) and reduce its laser power. Sweeping the pump frequency reveals a narrow electromagnetically induced transparency (EIT) peak within the broad single-photon depletion spectrum (see figure \ref{fig:Xspec}(c)), heralding the formation of a dark state necessary for STIRAP transfer. At bias magnetic fields of $<1$ G the Zeeman sub-levels of $(1)0_u^+(11,1)$ are essentially unresolved due to the $\sim$5 MHz excited state linewidth, consistent with the expected magnitude of the Zeeman shift \cite{McGuyer2013}. From measurements of Autler-Townes splittings we obtain $\Omega_2$, the angular Rabi frequency of the anti-Stokes coupling for various anti-Stokes laser powers. From a conservative estimate of the anti-Stokes laser beam waist, we find a transition strength of $8.6(9)\times 10^{-2} (ea_0)^2$ for ${X(0,0)\rightarrow (1)0_u^+(11,1)}$. This is one of the strongest molecular transitions in $^{88}\rm{Sr}_2$, with a value approaching that of a typical atomic transition.

\begin{table}
\caption{\label{tab:Xspecconst} Spectroscopic constants for $X^1\Sigma_g^+$ in units of $\rm{cm}^{-1}$. The binding energies were determined to an accuracy of 0.002 $\rm{cm}^{-1}$, limited by that of the wavelength meter (High Finesse WS7-60) used to determine the laser frequencies. The rotational splittings have tighter uncertainties because the \textit{relative} precision of the wavemeter is approximately $1\times 10^{-4}\rm{cm}^{-1}$. Only the first 5 vibrational levels were used, and the reduced $\chi^2$ of the fits to the binding energies and rotational splittings are 0.96 and 1.3, respectively.}
\begin{center}
\begin{tabular}{ccc}
\br
Spectroscopic constant& This work & References \cite{Stein2008,Stein2010}\\
\mr
$D_e$&1081.6436(20)&1081.64(2)\\
$\omega_e$&40.325(2)&40.328\\
$\omega_ex_e$&0.3986(4)&0.3994\\
$B_e$&0.017632(17)&0.01758\\
$\alpha_e$&$1.78(6)\times 10^{-4}$&$1.68\times 10^{-4}$\\
\br
\end{tabular}
\end{center}
\end{table}

\section{STIRAP}

\subsection{Numerical model}

We model the dynamics of the 5-level $\Lambda$-system (figure \ref{fig:Xspec}(d)) with the Lindblad master equation. Here, $|1\rangle,|2\rangle,|3\rangle,|4\rangle,|5\rangle$ represent the states $X(v=62,J=0,m_J=0)$, $X(0,0,0)$, $(1)0_u^+(11,1,-1)$, $(1)0_u^+(11,1,0)$, $(1)0_u^+(11,1,+1)$ respectively, and $m_J$ is the projection of the total angular momentum onto the lab-frame quantization axis $\hat{Z}$. The pump and anti-Stokes electric fields are $\vec{E}_j = \frac{\mathcal{E}_j}{2}\left(\vec{\epsilon}_{(j)}\,e^{-i\omega_j t}+\vec{\epsilon^*}_{(j)}\,e^{+i\omega_j t}\right)$ where $\omega_j$ are the laser angular frequencies and $j=\{1,2\}$ label the ground states. We pick the convention that the lasers propagate along the positive $\hat{Y}$ direction and write the polarization vectors as \begin{equation}
    \vec{\epsilon}_{(j)} = \hat{Z} \cos\theta_{j} + \hat{X} e^{i\phi_{j}}\sin\theta_{j},
\end{equation} where the inclination $\theta$ with respect to $\hat{Z}$ and the phase $\phi$ are angles that parametrize the polarization state. The angular Rabi frequencies coupling the ground states to the excited Zeeman sub-levels ($k=\{3,4,5\}$) are $\Omega_{j,k} \equiv \langle k|\vec{d}\cdot\vec{E}_j|j\rangle/\hbar$ where $\vec{d}$ is the dipole moment operator. Since the Clebsch-Gordan coefficients for $J=0\leftrightarrow 1$ are independent of $m_J$, we can write $\Omega_{j,k} = \Omega_j \epsilon_{(j),q}$ where \begin{equation}
    \epsilon_{(j),0} = \cos\theta_j,\, \epsilon_{(j),\pm1} = \mp\frac{1}{\sqrt{2}} e^{i\phi_j}\sin\theta_j.
\end{equation} Selection rules force $q=-1,0,+1$ for state labels $k=3,4,5$ respectively. This is true for both $j=1,2$ since we have a $\Lambda$-system. To make contact with the experiment, we note that the excitation rates and Autler-Townes splittings are both proportional to $\sum_q |\Omega_j \epsilon_{(j),q}|^2 = \Omega_j^2$ which are in turn proportional to the respective laser intensities, so the quantity that we measure in the preceding section is $\Omega_j$ even if the polarization is elliptical. Physically, this quantity is equivalent to the Rabi angular frequency for the ideal case of a $\pi$-transition ($J=0\leftrightarrow 1$, $\Delta m_J =0$) driven with pure linear polarization exactly parallel to the quantization axis.

In the electric dipole and rotating-wave approximation, the Hamiltonian governing the unitary evolution is \cite{vitanov1999adiabatic} \begin{align}
 H = 
   \begin{pmatrix}
  \delta_1 & 0 & \Omega_1\epsilon_{(1),-1}/2& \Omega_1\epsilon_{(1),0}/2& \Omega_1\epsilon_{(1),+1}/2& 0\\
  0 &\delta_2 & \Omega_2\epsilon_{(2),-1}/2& \Omega_2\epsilon_{(2),0}/2& \Omega_2\epsilon_{(2),+1}/2&0\\
  (\Omega_1\epsilon_{(1),-1})^*/2 &(\Omega_2\epsilon_{(2),-1})^*/2 & -Z& 0& 0 & 0\\
  (\Omega_1\epsilon_{(1),0})^*/2 &(\Omega_2\epsilon_{(2),0})^*/2 & 0& 0& 0 & 0\\
  (\Omega_1\epsilon_{(1),+1})^*/2 &(\Omega_2\epsilon_{(2),+1})^*/2 & 0& 0& +Z & 0\\
  0 &0 & 0& 0& 0 & 0\\
 \end{pmatrix},
\end{align} where $Z$ is the Zeeman splitting of the excited state, $\delta_j \equiv \omega_j - \omega_0$ are the angular frequency detunings from the $m_J=0$ sub-level with transition frequency $\omega_0$, and we include an additional auxiliary state $|6\rangle$ that does not participate in the coherent dynamics but merely functions to collect population decay from $|k\rangle$.  

To include the relaxation dynamics, we compute \begin{equation}
    \mathcal{L}_k(\rho) = -\frac{1}{2}\{G_k^\dag G_k,\rho \} + G_k\rho G_k^\dag,
\end{equation} where \{,\} is the anti-commutator, $G_k \equiv \sqrt{\Gamma}|k\rangle\langle 6|$ are the so called \textit{jump operators}, $\Gamma$ is the excited state linewidth, and $\rho$ is the $6\times 6$ density matrix for the whole system. The effect of $\mathcal{L}_k$ is to generate decay terms proportional to $-\Gamma$ in the diagonals of $|k\rangle$ and $-\Gamma/2$ in the off-diagonals between $|k\rangle$ and $|6\rangle$.

In some cases it is useful to include a phenomelogical decoherence rate $\Gamma_{\mathrm{eff}}$ which can be interpreted as the relative linewidth between the pump and anti-Stokes. To do so we compute \begin{equation}
    \mathcal{D}_{1,2}(\rho) = -\frac{\Gamma_{\mathrm{eff}}}{2} \left(P_1\rho P_2 + P_2\rho P_1 \right),
\end{equation} where $P_j \equiv |j\rangle\langle j|$ are projection operators onto the corresponding diagonal element. The effect of $\mathcal{D}_{1,2}$ is to generate decay terms proportional to $-\Gamma_{\mathrm{eff}}/2$ in the off-diagonals between $|1\rangle$ and $|2\rangle$.

The Lindblad master equation for our system is thus \begin{equation}\label{eq:lind}
    \frac{d}{dt} \rho = -i[H,\rho] +\mathcal{D}_{1,2}(\rho) +\sum_{k=3,4,5} \mathcal{L}_k(\rho),
\end{equation} where [,] is the commutator. We numerically solve the time evolution of equation \ref{eq:lind} for the input parameters $\Omega_j$, $\delta_j$, $\theta_j$, $\phi_j$, $Z$, $\Gamma$, and $\Gamma_{\mathrm{eff}}$, with the initial condition $\rho_{11}(t=0) =1$ and zero for all other entries. For later convenience, we define the one-photon (common) detuning $\Delta \equiv \delta_2$, and the two-photon (Raman) detuning $\delta \equiv \delta_1 - \delta_2$.

In the weak probe limit ($\Omega_1 \ll \Omega_2$) such that $\rho_{11} \approx 1$, $\rho_{22} \approx \rho_{33} \approx 0$, we can derive a general analytical expression for the excitation or EIT lineshape. Setting $Z\approx 0$ (unresolved excited states) and discarding negligible terms $O(\Omega_1/\Omega_2)$, we obtain \begin{align}
 \dot{\rho}_{11} \approx &\, -\sum_k\mathrm{Im}\left(\Omega^*_{1,k}\rho_{1k}\right),  \\ 
  \dot{\rho}_{1k} \approx &\, \left(-\frac{\Gamma}{2}-i\delta_1\right)\rho_{1k} + i \frac{\Omega_{1,k}}{2}+ i\frac{\Omega_{2,k}}{2}\rho_{12} \approx 0,\nonumber \\
  \dot{\rho}_{12} \approx &\, \left(-\frac{\Gamma_{\mathrm{eff}}}{2}-i\delta\right) \rho_{12} +i\sum_k\frac{\Omega_{2,k}^*}{2}\rho_{1k} \nonumber \approx 0.
\end{align} From the first equation, we see that the rate of excitation out of $|1\rangle$ is $R=\sum_k\mathrm{Im}\left(\Omega^*_{1,k}\rho_{1k}\right)$, where $\mathrm{Im}()$ denotes the imaginary part. To find the excitation rate approaching steady-state conditions, we set $\dot{\rho}_{12} \approx \dot{\rho}_{1k}  \approx 0$ to solve for $\rho_{1k}$ and substituting back into the expression for $R$, we find \begin{align} \label{eq:EITline}
    R = &\frac{\Gamma}{\Gamma^2 + 4\delta_1^2} \\ &\times\left[\Omega_1^2 - \left(
    \sum_k\Omega_{1,k}^*\Omega_{2,k}\right)\left(\sum_k\Omega_{1,k}\Omega^*_{2,k}\right) \frac{\Omega_2^2-8\delta\delta_1+\Gamma_{\mathrm{eff}}\Gamma(1-4\delta_1^2/\Gamma^2)}{|\Omega_{2}^2+(\Gamma+2i\delta_1)(\Gamma_{\mathrm{eff}}+2i\delta)|^2}\right]. \nonumber
\end{align} The general EIT lineshape for non-cycling transitions is then \begin{equation}
    N(\delta,
\delta_1,t) = N_0\, \exp\left[-Rt\right]. \label{eq:generalrate}
\end{equation} 

Finally, for the ideal case where the Raman lasers have exactly the same polarization (i.e. $\theta_1 = \theta_2$, $\phi_1 =\phi_2$), we have $\sum_k\Omega_{1,k}^*\Omega_{2,k} = \sum_k\Omega_{1,k}\Omega^*_{2,k} = \Omega_1\Omega_2$ since both ground states have $J=0$, and equations \ref{eq:EITline} and \ref{eq:generalrate} further simplify to \begin{equation}
    N(\delta,
\delta_1,t) = N_0 \exp\left[-t\,\frac{\Gamma\Omega_{1}^2}{\Gamma^2+4\delta_1^2} \left(1-\Omega_2^2\,\frac{\Omega_2^2-8\delta\delta_1+\Gamma_{\mathrm{eff}}\Gamma(1-4\delta_1^2/\Gamma^2)}{|\Omega_{2}^2+(\Gamma+2i\delta_1)(\Gamma_{\mathrm{eff}}+2i\delta)|^2}\right) \right], \label{eq:eitfit}
\end{equation} where the single-photon and two-photon effects are separated into distinct terms; e.g. for $\Omega_2 = 0$, we immediately recover equation \ref{eq:depletion}. Equation \ref{eq:eitfit} can be shown to be in exact agreement with reference \cite{debatin2011molecular}, and is also valid in the case where the excited state structure is very well separated such that one effectively addresses a three-level $\Lambda$-system. As can be seen in figure \ref{fig:Xspec}(c), the analytical form of equation \ref{eq:eitfit} is an excellent fit to the experimental data.

\subsection{STIRAP in a non-magic optical lattice}

\begin{figure}
    \centering
    \includegraphics[width=\textwidth]{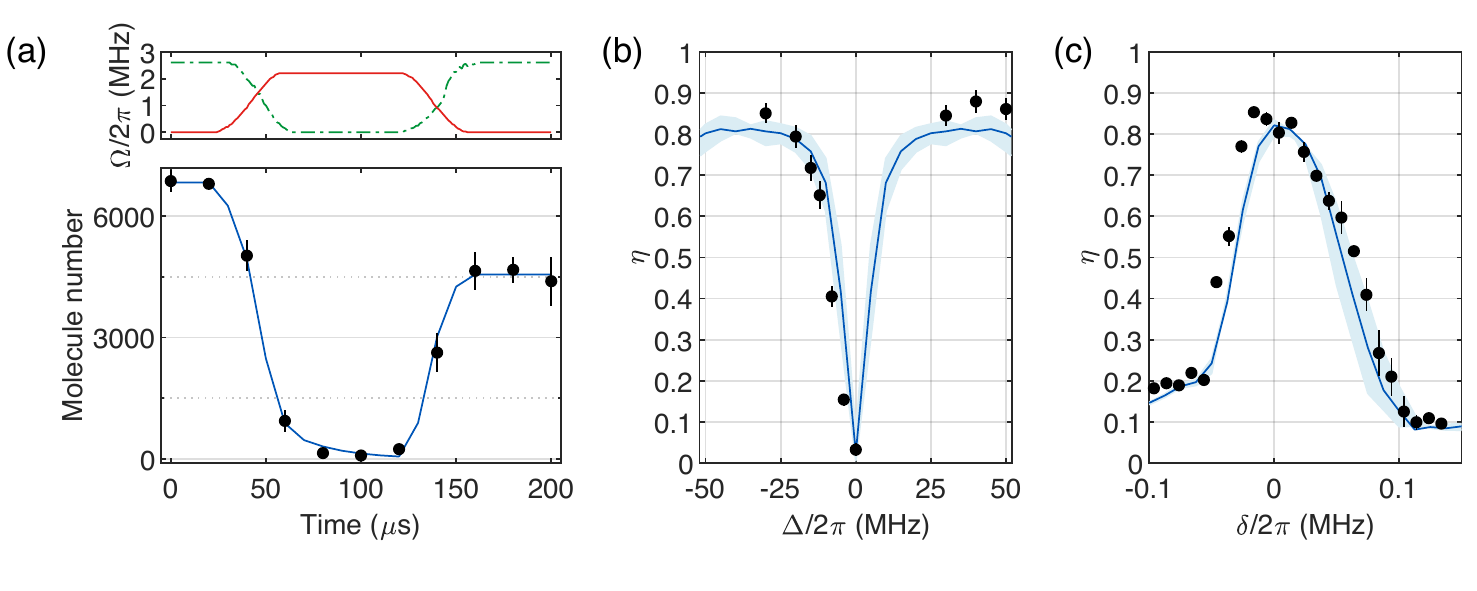}
    \caption{Creation of $\rm{Sr}_2$ dimers in the absolute ground state. (a) Molecules initially in $X(62,0)$ are transferred using STIRAP to the absolute ground state $X(0,0)$ within $40 \,\mu s$. To detect the transferred molecules we reverse the order of the STIRAP pulses to recover the initial state. Throughout the roundtrip ($200 \,\mu \rm{s}$), the optical lattice trap is switched off and the molecules are in free flight. Solid blue line is the model prediction with no free parameters; i.e. we independently measured the time evolution of $\Omega_1$ and $\Omega_2$ (solid red and dashed-dotted green lines respectively), a relative polarization angle of 16(2) degrees, and use $\Gamma_{\mathrm{eff}} = 2\pi\times\, 3.5(5) \,\rm{kHz}$. For this trace, $\Delta/2\pi = +30 \,\rm{MHz}$. (b) One-way transfer efficiency versus the common (one-photon) detuning, $\Delta$. The efficiency $\eta$ drops near resonance because of scattering arising from experimental imperfections in controlling laser polarization. (c) One-way transfer efficiency versus the Raman (two-photon) detuning, $\delta$. In both (b) and (c), the solid blue curve is the model prediction, and the light-blue shaded area covers the range of simulation results given the uncertainty of the measured parameters. All error bars represent $1\sigma$ of statistical error.}
    \label{fig:STIRAP}
\end{figure}

Figure \ref{fig:STIRAP}(a) shows a representative time evolution of the number of $X(62,0)$ molecules during a roundtrip STIRAP at a common detuning of $\Delta = 2\pi \times 30 \,\rm{MHz}$, and angular Rabi frequencies $\Omega_1 = 2\pi\times 2.2\,\rm{MHz}$ and $\Omega_2=2\pi\times 2.6\,\rm{MHz}$. Here the lattice is tuned to a wavelength of $\lambda = 914.0(1)\,\rm{nm}$, and we perform STIRAP in free flight by switching the lattice trap off for the entirety of the roundtrip to eliminate lattice induced thermal decoherence due to the polarizability difference of $X(62,0)$ and $X(0,0)$. Auxiliary measurements of the molecular cloud size indicate that switching off the trap on timescales of $<200\, \mu s$ does not result in significant heating or number loss. Moreover, the first order Doppler broadening is expected to be manageable at the level of $f_0 \sqrt{3k_BT/M}/c \sim 3.5(5)\,\rm{kHz}$ at temperatures of $T = 8(2) \mu \rm{K}$ for a Raman transition frequency of $f_0 \approx 31.825\,\mathrm{THz}$; $M$ is the molecular mass. The roundtrip transfer efficiency is quantified as $\eta^2 = (N_3-N_2)/N_1$, where $N_1, N_3$ are the initial and final molecule numbers respectively, and $N_2$ is the remaining molecule number after the forward transfer. For our Rabi frequencies, we typically achieve full extinction such that $N_2 =0$, except in a few extreme cases where $\Delta$ is very large resulting in reduced adiabaticity. Assuming equal efficiencies for the forward and reverse transfer, we routinely achieve one-way transfer efficiencies of $\eta = 85(3)\%$.

Due to the unresolved Zeeman structure of $(1)0_u^+(11,1)$, a finite relative angle between the polarizations of the Raman (pump and anti-Stokes) lasers dilutes the Rabi couplings and destabilizes the dark state. This leads to increased near-resonant scattering which diminishes the overall transfer efficiencies. In the current study, we measure a relative polarization angle of 16(2) degrees between the pump and anti-Stokes. The cause was traced to a dichroic mirror combining the lasers with the lattice. While we could solve the issue by placing a Glan-Thompson polarizer after this dichroic mirror and immediately before the chamber viewport, geometric constrains in our current setup prohibit this. Nevertheless, for $J=0\leftrightarrow 1$ transitions, we can circumvent this by either lifting the degeneracy of Zeeman sublevels, or performing STIRAP at $\Delta$ larger than $\Gamma$, the linewidth of the ${X\rightarrow 0_u^+}$ transition, as demonstrated in figure \ref{fig:STIRAP}(b). Numerical simulations using the measured relative polarization angle and expected $\Gamma_{\mathrm{eff}}$ show good agreement. We note that detuned STIRAP has been reported to mitigate other technical imperfections such as laser phase noise and stray reflections \cite{seeselberg2018,panda2016,yatsenko2014detrimental}. When choosing an operational common detuning, it is prudent to carefully survey the molecular structure. For instance, in $^{88}\rm{Sr}_2$, the rotational splitting of $X(62,2)$ and $X(62,0)$ is approximately 70 MHz. Therefore, blue detuning is preferred so as to avoid accidental perturbation of the excited state by the pump laser should it be tuned close to $0_u^+(11,1) \rightarrow X(62,2)$.  Similarly, technical leakage light through the acousto-optic modulators used to modulate the laser intensities can diminish transfer efficiencies should the residual diffraction orders accidentally address a $X\rightarrow 0_u^+$ resonance. These technical effects are non-negligible since the laser intensities used in STIRAP are large.

A hallmark of STIRAP that gives it an edge over other transfer schemes (e.g. a Raman $\pi$-pulse \cite{Kondov2019,Cairncross2021}) is its robustness against small perturbations (e.g. laser intensity stability, frequency drifts) as evident by the detuned-STIRAP resonance in figure \ref{fig:STIRAP}(c), where the efficiency remains ${>50}\%$ even as the relative Raman detuning is scanned over ${>100}\, \rm{kHz}$. This range can be made wider with larger laser intensities. The detuned-STIRAP lineshape also shows the expected asymmetry with the sharper edge facing the one-photon resonance location.

\subsection{STIRAP in a magic lattice}

\begin{figure}
    \centering
    \includegraphics[width=\textwidth]{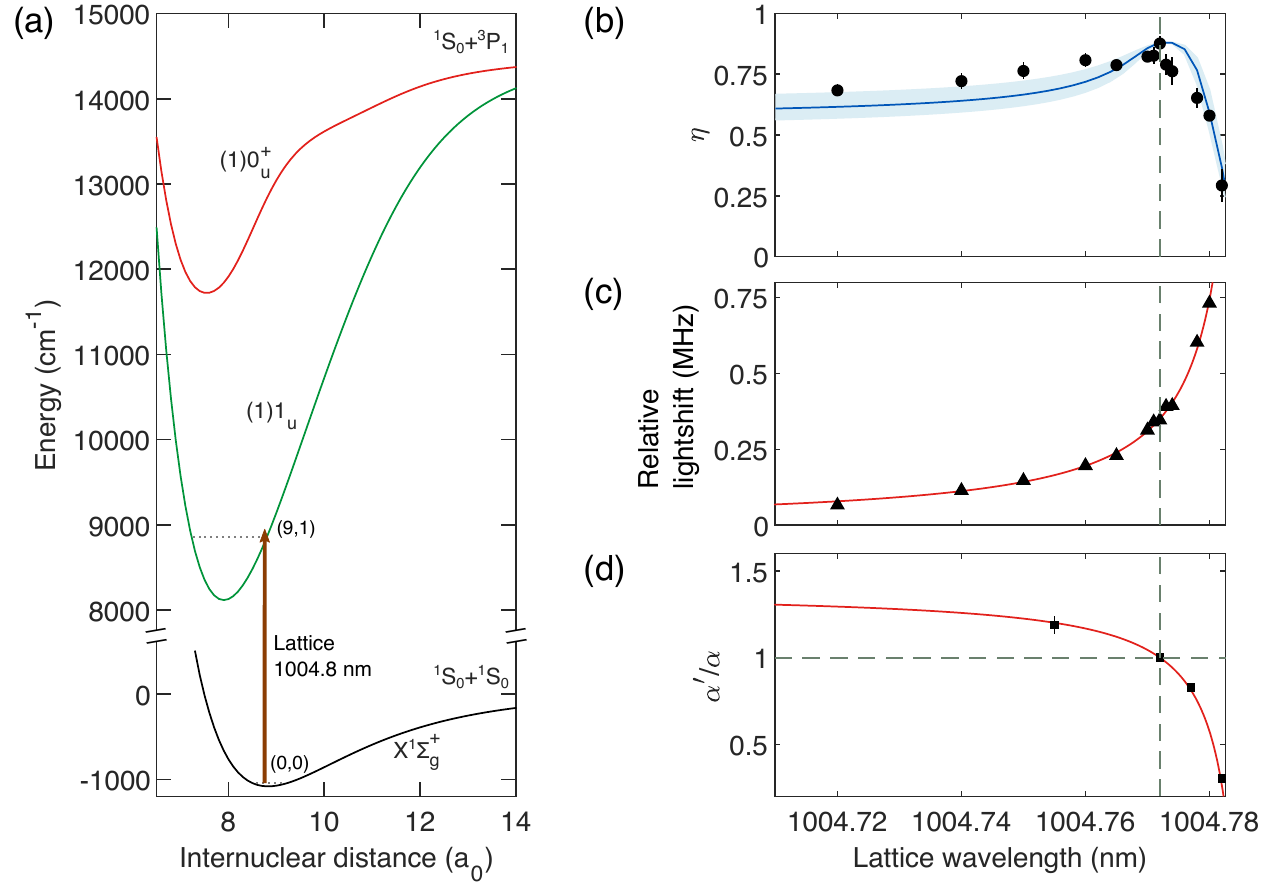}
    \caption{STIRAP transfer in a magic-wavelength optical lattice. (a) By tuning the optical lattice near the $(1)1_u(9,1)$, we engineer a magic trap for the pair $X(62,0)$ and $X(0,0)$ where their polarizabilities are equal. (b) Enhancement of STIRAP efficiency at the magic wavelength:  measured (black circles) and modeled (solid blue) with the range of simulation results (shaded light-blue area) for $T=8(2)\,\mu\rm{K}$. (c) Corresponding lattice-induced shift of the STIRAP resonance. (d) Polarizability ratio of $X(0,0)$ to $X(62,0)$ around the magic wavelength, determined from the differential lattice lightshifts. In both (c) and (d), the red solid line is the fit to the data in the form $a/(x-x_0)+b$ where $a$ and $b$ are free parameters and $x_0$ is fixed to the independently measured $X(0,0)\rightarrow (1)1_u(9,1)$ resonance wavelength. All error bars represent $1\sigma$ of statistical error.}
    \label{fig:magic}
\end{figure}

We now explore performing STIRAP in a deep optical lattice, at a trap depth of $U_0 = 1009(44) \,E_{\mathrm{rec}}$. Unlike in the previous subsection, here we leave the trap light on throughout the sequence. Our current strategy for engineering magic lattices involves a lattice blue-detuned from a transition connecting the deeply bound ground state with a narrow rovibronic state in the $(1)1_u$ potential (ungerade, $\Omega=1$) that asymptotes to the ${^1S}_0+{^3P}_1$ threshold \cite{leung2020,Kondov2019}, as illustrated in figure \ref{fig:magic}(a). At the magic wavelength, the polarizabilities of the two $J=0$ vibrational ground states are matched ($\alpha^\prime/\alpha = 1$), resulting in equal trap depths. This effectively removes lattice-induced thermal decoherence and precludes the excitation of breathing modes during the state transfer.

We observe enhancement of STIRAP efficiency in a magic wavelength lattice as shown in figure \ref{fig:magic}(b), accompanied by lattice-induced Stark shift of the peak STIRAP efficiency (figure \ref{fig:magic}(c)). To understand this, we measure the polarizability ratio of the ground states at various lattice wavelengths near the ${X\rightarrow 1_u}$ transition (figure \ref{fig:magic}(d)),  \begin{equation}\label{eq:polratio}
    \frac{\alpha^\prime}{\alpha} = 1-4f_{\mathrm{rec}}\,\frac{L_0}{f_{\mathrm{ax}}^2},
\end{equation} where $\alpha^\prime, \,\alpha$ are the scalar polarizabilities of $X(0,0)$ and $X(62,0)$ respectively, $L_0$ is the differential lattice-induced lightshift, $f_{\mathrm{ax}}$ is the axial trap frequency (section 5), and $f_{\mathrm{rec}} \equiv h/(2M\lambda^2) = E_{\mathrm{rec}}/h$ is the standard expression for the recoil frequency. Since $L_0$ and $f_{\mathrm{ax}}^2$ both depend linearly on lattice intensity, this calculation bypasses the need to determine any geometric parameters (e.g. lattice beam waist).

For $X(0,0)$, the optimal magic wavelength \cite{leung2020} with the greatest magic detuning ($\Delta_m = 4.456(3) \,\rm{GHz}$) occurs at 1004.7723(1) nm, blue detuned from $X(0,0) \rightarrow (1)1_u(9,1)$ with a measured transition strength of $1.33(15) \times 10^{-4} \, (ea_0)^2$ using an all-frequency method following reference \cite{leung2020}. This is theoretically predicted to be the strongest ${X\rightarrow 1_u}$ transition that is below the $(1)0_u^+$ potential minimum in $^{88}\rm{Sr}_2$. In the Lamb-Dicke regime, the first-order Doppler effect is suppressed ($\Gamma_{\mathrm{eff}} = 0$). In general, the lattice-induced differential lightshift on a transition in a non-magic trap becomes compounded for higher trap motional states. For an experimentally set (bare) Raman detuning of the pump and anti-Stokes in the motional ground state, the thermal distribution of the molecules occupying various trap motional states maps to a probability density, $p$, for the molecule to experience a Raman detuning \textit{additionally} shifted by $-\delta^\prime$  \begin{equation}p(-\delta^\prime) =\begin{cases}
      \frac{1}{2}(B\delta^\prime)^2 e^{-B\delta^\prime}, & B\delta^\prime\geq 0 \\
      0, & B\delta^\prime< 0 \label{eq:carriertherm}
\end{cases}\end{equation} where $B\equiv \frac{h}{k_BT\left(\sqrt{\alpha^\prime/\alpha}\, - 1\right)}$ is a factor that depends on the temperature and polarizability mismatch \cite{mcdonaldzelevinsky2015}. This probability density $p(-\delta^\prime)$ peaks at $\delta^\prime_{max} = 2/B$. To model the transfer efficiencies, we first simulate ideal STIRAP efficiencies, $\eta(\delta)$, for the measured parameters as a function of $\delta$. The function $\eta(\delta)$ peaks at approximately $\delta = 0$. Thus, conceptually the overall efficiency involves an overlap integral between $\eta$ and $p$, which we maximize in the actual experiment by manually adjusting the bare Raman detuning (as in figure \ref{fig:magic}(c)). We can account for this experimental detail to some extent in the model by shifting $\delta^\prime \rightarrow \delta^\prime -\delta^\prime_{max}$ such that the peaks of the distributions $p$ and $\eta$ line up. The overall efficiency is then calculated as the thermal average of $\eta$ (i.e. a convolution) \begin{equation}
    \langle \eta \rangle  = \frac{\int_{-\infty}^\infty d\delta\: \eta(\delta)p(\delta -\delta^\prime_{max})}{\int_{-\infty}^\infty d\delta \: p(\delta)},
\end{equation} and the lattice wavelength dependence enters implicitly through $\alpha^\prime/\alpha$. The solid line in figure \ref{fig:magic}(b) shows the result of the simulation which reproduces the salient features of the measurement fairly well.  

\section{Two-body ultracold reactive collisions \label{sec:loss}}

\begin{figure}
    \centering
    \includegraphics[width=0.8\textwidth]{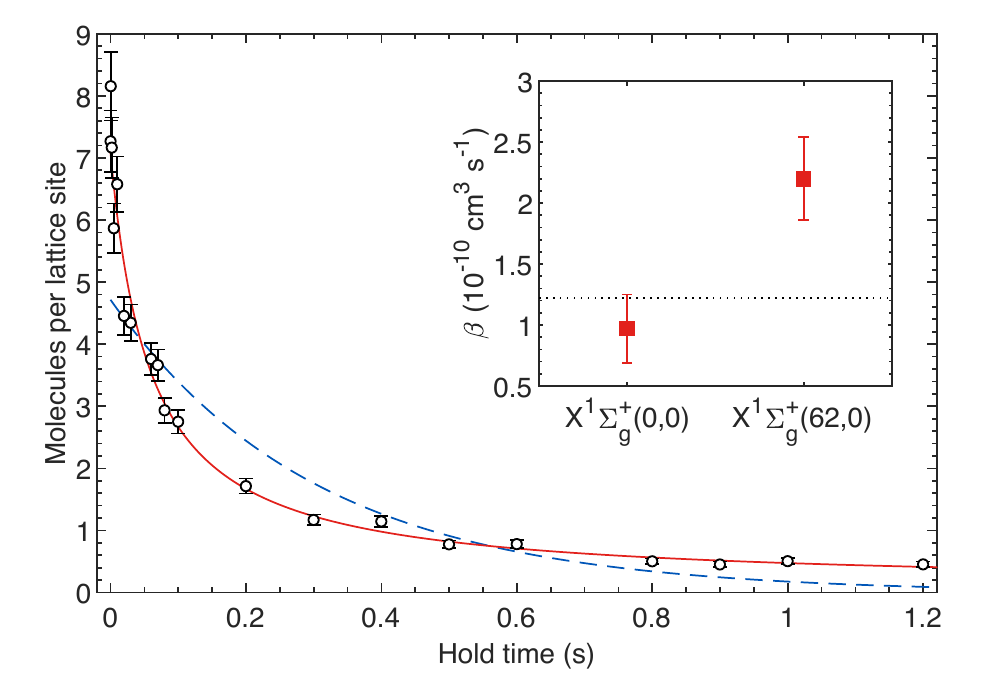}
    \caption{Collisional losses of ${X^1\Sigma}_g^+$ molecules.  The molecule number ($X(0,0)$, black circles) decays over the hold time in the optical lattice.
    Error bars represent $1\sigma$ of systematic error in determining the molecule number per lattice site. Fits to the rate equation $\dot{N} = -k N^\gamma$ with $\gamma=1$ (dashed blue line) and $\gamma=2$ (solid red line) suggest two-body loss. Inset: Two-body loss rates for $J=0$ molecules in the lowest and least bound vibrational states in ${X^1\Sigma}_g^+$ are found to be close to the universal limit (black dotted line). The experimental values are averaged over more than 10 runs. Error bars here represent $1\sigma$ of error arising from statistical fluctuations and the systematic uncertainty from density calibration.}
    \label{fig:loss}
\end{figure}

With the creation of large numbers of molecules in ${X^1\Sigma}_g^+(0,0)$, we are in a good position to study state-specific losses of an ultracold gas of alkaline-earth-metal dimers. To this end, we prepare a purified sample of $J=0$ molecules by wiping away $J=2$ from the initial photoassociated mixture (see section 2). We hold the $v=0$ ground state molecules after a forward STIRAP sequence in a non-magic optical lattice ($\lambda = 914.0(1)$ nm) for a variable amount of time, and then reverse the STIRAP sequence to recover weakly bound molecules which we detect. Figure \ref{fig:loss} shows the decay of the $v=0$ trapped molecule number over time. Fitted curves to the rate equation $\dot{N} = -k N^\gamma$ (where $k$ is a free parameter) strongly suggest two-body collisions ($\gamma =2$) to be the dominant loss channel. 

We can extract the two-body loss parameter, $\beta$, with a density calibration. Following reference \cite{Guo2018prx}, the rate equation can be written as \begin{equation}
    \dot{N}(t) =  - \beta \frac{A}{T^{3/2}} N(t)^2,
\end{equation} where $N$ is the number of molecules per lattice `pancake', $A \equiv \left(\bar{\omega}^2 M/4\pi k_B\right)^{3/2}$, and $\bar{\omega} \equiv 2\pi (f_{\mathrm{ax}} f_{\mathrm{rad}}^2)^{1/3}$. To simplify the analysis, in this study we assume that the molecules remain at the same temperature, $T$, throughout the hold duration. 

\begin{figure}
    \centering
    \includegraphics[width=\textwidth]{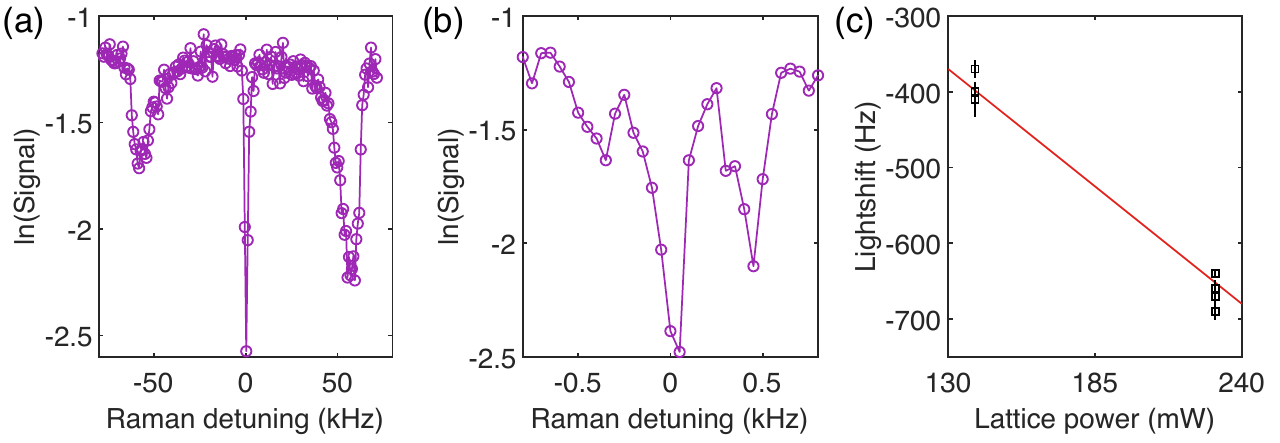}
    \caption{Lattice trap frequencies measured spectroscopically via the naturally narrow, shallow-to-shallow Raman transition ${X(62,0)\rightarrow X(61,0)}$. Here the Raman probes are counter-propagating along the lattice axis and blue-detuned by 1.5 GHz from $(1)0_u^+(12,1)$. Scanning the pump reveals the (a) axial and (b) radial sidebands. Purple solid lines are guides to the eye. (c) Small differential lightshift of ${X(62,0)\rightarrow X(61,0)}$ versus lattice power. This can be converted to a polarizability ratio $\alpha_{v=61}/\alpha_{v=62} = 1.000932(29)$ at a trap wavelength of $914.0(1)$ nm. All error bars represent $1\sigma$ of statistical error.}
    \label{fig:sidebands}
\end{figure}

\begin{figure}
    \centering
    \includegraphics[width=\textwidth]{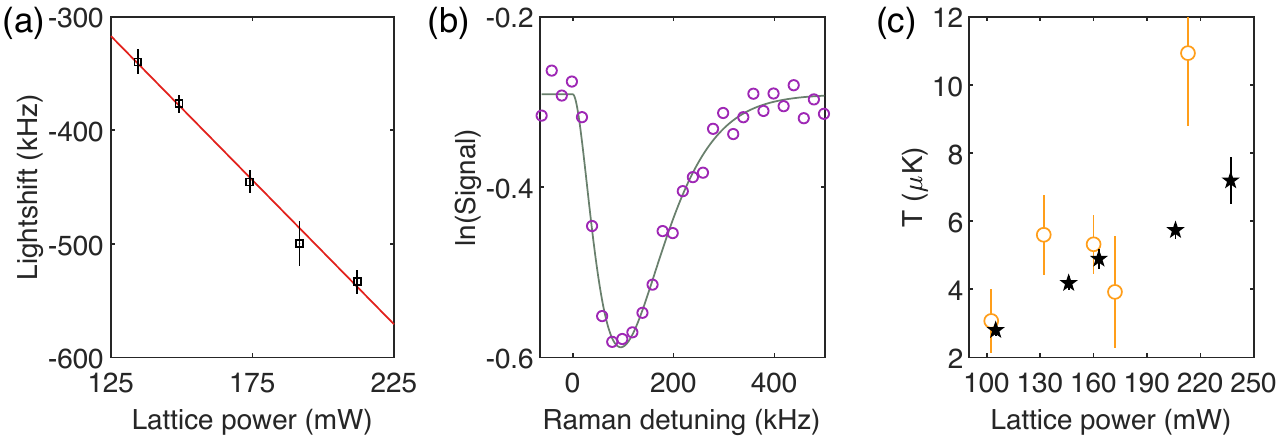}
    \caption{Raman carrier thermometry of lattice-trapped molecules. (a) Differential lightshift (black squares) of ${X(62,0)\rightarrow X(0,0)}$. Together with the trap frequency measurements, these imply a polarizability ratio $\alpha^\prime/\alpha =  1.5176(11)$ at a trap wavelength of $914.0(1)$ nm, allowing for the measured trap frequencies of $X(62,0)$ to be scaled to that of $X(0,0)$. Solid red line is a linear fit to the data. (b) Shallow-to-deep Raman transition ${X(62,0)\rightarrow X(0,0)}$ via $(1)0_u^+(11,1)$ in a non-magic lattice. The thermal distribution of the molecules is imprinted onto the carrier via lightshifts. Solid line is the fit to the data in the form of equation \ref{eq:carriertherm}. Here, the probes are co-propagating and red-detuned from the intermediate state by 3 GHz.  (c) Molecular temperature determined from the thermal broadening of the carrier (black stars). Auxiliary temperature measurements from the ratio of the axial sideband areas (orange circles) are consistent with the carrier method. All error bars represent $1\sigma$ of statistical error.}
    \label{fig:densitycal}
\end{figure}

Since the molecules are tightly trapped in the Lamb-Dicke and resolved sideband regimes, we can spectroscopically access the axial and radial trap frequencies ($f_{\mathrm{ax}}$ and $f_{\mathrm{rad}}$ respectively) by probing the shallow-to-shallow Raman transition ${X(62,0)\rightarrow X(61,0)}$ via $(1)0_u^+(12,1)$ as shown in figures \ref{fig:sidebands}(a) and (b). Here, we use \textit{counter-propagating} probe beams to maximize the imparted momentum along the axial lattice direction so as to observe the axial sidebands. This Raman transition is nearly magic since the polarizability difference of adjacently bound near-threshold states is negligible (figure \ref{fig:sidebands}(c)). As $\bar{\omega}^2$ is measured for $X(62,0)$, in the case of $X(0,0)$ we must further scale $\bar{\omega}^2$ by the polarizability ratio $\alpha^\prime/\alpha$. From measurements of differential lightshift for the shallow-to-deep Raman transition ${X(62,0)\rightarrow X(0,0)}$ (figure \ref{fig:densitycal}(a)) and equation \ref{eq:polratio}, we find $\alpha^\prime/\alpha= 1.5176(59)$ at $\lambda = 914.0(1)$ nm. 

The next step is to determine the temperature of the molecular ensemble. We do so using carrier thermometry \cite{mcdonaldzelevinsky2015} with the shallow-to-deep Raman transition ${X(62,0)\rightarrow X(0,0)}$ and \textit{co-propagating} probes (see figure \ref{fig:densitycal}(b)). For the purposes of inferring temperature via thermally imprinted lattice lightshifts (i.e. equation \ref{eq:carriertherm}), probing a transition with a large polarizability difference is advantageous and has better accuracy since the effect is exaggerated over other broadening sources (e.g. power broadening) \cite{han2018carrier}. In figure \ref{fig:densitycal}(c) we compare the temperatures extracted using carrier thermometry to the conventional method of taking the ratio of the integrated area under the sidebands and find good agreement.

The measured two-body loss rate coefficient $\beta$ for $X(0,0)$ and $X(62,0)$ are $0.97(28) \times 10^{-10}\, \mathrm{cm^3\,s^{-1}}$ and $2.2(0.3) \times 10^{-10}\, \mathrm{cm^3\,s^{-1}}$ respectively, which are found to be unaffected by the presence of $J=2$ vibrationally least bound molecules. These results are summarized in the inset to figure \ref{fig:loss}. An estimate of the universal inelastic loss rate via the dominant $s$-wave channel for $X(0,0)$ yield $1.22 \times 10^{-10}\, \mathrm{cm^3\,s^{-1}}$, independent of the ensemble temperature \cite{Idziaszek2010prl}. This estimate relies on a coupled cluster computation for the isotropic van der Waals coefficient, $C_6$, using the explicitly connected representation of the expectation value and polarization propagator \cite{jeziorski1993explicitly,moszynski2005time} and the best approximation XCCSD4 method \cite{korona2006time}. We find $C_6= 15685 \,\rm{a.u.}$ for $\rm{Sr}_2$ dimers at the equilibrium distance $R_e = 8.829 \,a_0$, consistent with an independent calculation from reference \cite{juliennePrivateComm}. We had previously applied the same method and basis set to calculate the leading van der Waals coefficient for atomic strontium in the ground state and found excellent agreement with high-resolution Fourier transform spectra \cite{Skomorowski2012jcp,Stein2010}, thus we expect a similar level of accuracy for the absolute ground state strontium molecules in the present investigation. The closeness of our measured $\beta$ for $X(0,0)$ molecules to the universal loss limit suggests that the molecules react with near unity probability following a collision at short range. For a homonuclear alkaline-earth dimer in the absolute ground state, one possible exoergic process that will manifest in such a loss is the formation of stable trimers \cite{juliennePrivateComm}. Another possibility is the photoexcitation of ``sticky'' four-body complexes by the intense lattice light \cite{christianen2019prl,Gregory2020prl}. For the near-threshold $X(62,0)$ state, the slightly larger loss rate may indicate that vibrational relaxation effects are non-negligible. The results, however, do not limit the prospects of using these molecules for precision spectroscopy, since the implementation of a 3D lattice can strongly suppress collisional losses and enable long interrogation times \cite{chotiaprl2012}.

\section{Conclusion}

In summary, we have created ultracold $^{88}\rm{Sr}_2$ molecules in the absolute ground state using STIRAP to coherently transfer photoassociated molecules. We achieve favorable transfer efficiencies of nearly 90\%, both in free flight and in a deep optical lattice tuned to a magic wavelength, limited by the available pump laser power. Future experiments should benefit from a redesigned experimental setup such that the relative polarization of the pump and anti-Stokes beams can be made more parallel. We have mapped out all 63 vibrational states with $J=0,2$ in $X^1\Sigma_g^+$ to an accuracy of 0.002 $\rm{cm}^{-1}$. We spectroscopically observed the effect of spin-orbit coupling on the bound states of $(1)0_u^+$, verified its potential depth, and identified a favorable STIRAP pathway from the measured transition strengths. We investigated the lifetime of absolute ground state molecules in an optical trap and find two-body collisions near the universal loss rate. 

The work presented here demonstrates the ability to access the full vibrational state space of $X^1\Sigma_g^+$ closed-shell molecules, and in doing so, opens up the variety of experiments where efficient state initialization in a deeply bound level is necessary. The $X\rightarrow 1_u$ transition described here presents one of the most favorable conditions for engineering a near-resonant magic lattice for Raman transitions within the ground potential, so as a natural next step we plan to utilize this to explore the stability and accuracy of the vibrational molecular clock. Large samples of ground state molecules should allow us to gain insight into the photon scattering process in a near-resonant magic lattice \cite{Kondov2019}, and enable loading into 3D magic lattices at mid-IR wavelengths tuned far below the minima of the excited potentials where we expect the main loss processes to be diminished. With these, tests of fundamental physics with alkaline-earth-metal molecules are now within reach.

\ack
We thank H. Bekker for early contributions to this work, S. Will for the loan of equipment, and M. Borkowski for critical reading of the manuscript.  This work was supported by NSF grant PHY-1911959, AFOSR MURI FA9550-21-1-0069, ONR grant N00014-17-1-2246, ONR DURIP N00014-20-1-2646, and a Center for Fundamental Physics grant from the John Templeton Foundation \& Northerwestern University.  R. M. acknowledges the Polish National Science Center Grant No. 2016/20/W/ST4/00314.

\appendix
\section{$(1)0_u^+$ binding energies and transition strengths}

Table \ref{tab:0ubindtable} shows the binding energies for the lowest 37 vibrational states of $(1)0_u^+$ with $J=1$ via laser excitation from $X(62,0)$, with respect to the atomic threshold ${^1S}_0+{^3P}_1$. We take the intercombination frequency from reference \cite{Ferrari2003} and the binding energy of $X(62,0)$ from reference \cite{McDonald2017}. 

For $v=9$-$20$ we use a frequency comb to determine the pump laser frequency with a high precision. The $X\rightarrow 0_u^+$ resonance frequencies are linearly extrapolated to zero lattice intensity, and we report uncertainties inflated by the square root of the reduced $\chi^2$ if it is greater than unity. The rest were determined using a wavelength meter to an uncertainty of 60 MHz. 

In addition, we also list $S_{X(62,0)}$ and $S_{X(61,0)}$, the measured $X\rightarrow 0_u^+$ transition strengths from $X(62,0)$ and $X(61,0)$ respectively. The method of determining these transition strengths is described in the main text.

\Table{\label{tab:0ubindtable} Binding energies for the lowest 37 vibrational states of $(1)0_u^+$ with $J=1$ with respect to the atomic threshold ${^1S}_0+{^3P}_1$. Also shown are $S_{X(62,0)}$ and $S_{X(61,0)}$, the measured $X\rightarrow 0_u^+$ transition strengths from $X(62,0)$ and $X(61,0)$ respectively, in units of $(10^{-5}(ea_0)^2)$. These values are plotted in figure \ref{fig:0uspec}. All reported values are obtained in this work.}
\br
$(1)0_u^+(v,J=1)$ & Binding energy (THz) & $S_{X(62,0)}$ & $S_{X(61,0)}$\\
\mr
0  & 82.462,408(60) & - & -  \\
1  & 80.056,883(60) & - & - \\
2  & 77.669,758(60) & - & - \\
3  & 75.301,073(60) & - & - \\
4  & 72.950,983(60) & - & - \\
5  & 70.619,778(60) & - & - \\
6  & 68.307,898(60) & - & - \\
7  & 66.016,088(60) & - & - \\
8  & 63.745,468(60) & - & - \\
9  & 61.497,833,938(122) & 1.15(24) & - \\
10 & 59.275,910,257(346) & 0.0797(84) & - \\
11 & 57.084,156,509(120) & 1.60(23) & 7.1(1) \\
12 & 54.929,909,904(110) & 0.242(20) & 1.01(11) \\
13 & 52.825,770,427(44) & 0.327(42) & - \\
14 & 50.791,292,559(96) & 0.300(16) & - \\
15 & 48.855,512,131(176) & 0.258(58) & - \\
16 & 47.036,183,941(226) & 0.147(17) & 0.630(67) \\
17 & 45.320,332,029(140) & 1.15(12) & 5.54(84) \\
18 & 43.686,692,338(92) & 0.0161(12) & 0.0873(92) \\
19 & 42.124,534,970(66) & 0.620(66) & 3.6(9) \\
20 & 40.655,031,206(80) & 0.274(39) & - \\
21 & 39.174,338(60) & - & - \\
22 & 37.763,228(60) & - & - \\
23 & 36.377,628(60) & - & - \\
24 & 35.066,118(60) & - & - \\
25 & 33.737,818(60) & - & - \\
26 & 32.397,458(60) & - & - \\
27 & 31.241,318(60) & - & - \\
28 & 29.959,413(60) & - & - \\
29 & 28.947,228(60) & - & - \\
30 & 27.621,073(60) & - & - \\
31 & 26.361,623(60) & - & - \\
32 & 25.400,438(60) & - & - \\
33 & 24.186,203(60) & - & - \\
34 & 23.295,548(60) & - & - \\
35 & 22.096,778(60) & - & - \\
36 & 21.284,598(60) & - & - \\
\br
\end{tabular}
\end{indented}
\end{table}

\clearpage

\section*{References}
\input{main.bbl}

\end{document}

%% file: main.bbl
\providecommand{\newblock}{}